\documentclass[onecolumn,11pt]{article}
\usepackage{setspace}
\usepackage{amsmath, amsthm, amssymb}
\usepackage[dvips]{graphicx}
\usepackage[round,sort]{natbib}
\usepackage{array}
\usepackage{color,tikz}
\usepackage{multirow}
\usepackage[Large,up,hang]{subfigure}
\usepackage[left=01in,top=1in,right=1in,bottom=1in,includefoot,nohead]{geometry}

\usepackage{latexsym}           
\usepackage{ifpdf}
\usepackage{enumerate}          
\usepackage{mathrsfs}           
\usepackage[plain]{algorithm}
\usepackage{algorithmic}
\usepackage{makeidx}            
\usepackage{array}
\usepackage{longtable}
\usepackage{geometry}           
\usepackage{verbatim}           

\RequirePackage[OT1]{fontenc}
\RequirePackage[colorlinks,citecolor=blue,urlcolor=blue]{hyperref}
\RequirePackage{hypernat}

\usepackage{textcomp}

\usepackage{amsmath}

\usepackage{hyperref}
\hypersetup{
    colorlinks,%
    citecolor=blue,%
    filecolor=black,%
    linkcolor=blue,%
    urlcolor=blue
}

\newcommand{\keywords}[1]{\par\noindent
{\small{\em Keywords\/}: #1}}

\begin{document}

\title{Hybrid-Lambda: simulation of multiple merger and Kingman gene genealogies in species networks and species trees}
\author{Sha Zhu\,$^{1,*}$, James H Degnan\,$^{2}$ and Bjarki Eldon\,$^3$\\
$^{1}${\it Wellcome Trust Centre for Human Genetics, University of Oxford, UK}\\
$^{2}${\it Biomathematics Research Centre, University of Canterbury,} \\{\it Christchurch, New Zealand}\\
 $^{3}${\it Technische Universit\"at Berlin, Germany}\\
$^*$to whom correspondence should be addressed}

 \maketitle


\begin{abstract}
  \indent {\tt Hybrid-Lambda} is a software package that simulates gene trees under Kingman or two Lambda-coalescent processes within species networks or species trees. It is written in {\tt C++}, and released under GNU General Public License (GPL) version 3. Users can modify and make new distribution under the terms of this license. For details of this license, visit {\tt http://www.gnu.org/licenses/}.  {\tt Hybrid-Lambda} is available at https://code.google.com/p/hybrid-lambda.\\
\keywords{hybridization, lambda coalescent, gene tree, phylogenetic network}
  \end{abstract}






\indent Species trees describe ancestral relations among species. Gene trees
describe the random ancestral relations of alleles sampled within species.  Gene trees and species trees are often
 assumed to be bifurcating
\citep{Degnan2005,Hudson1990,K82}.  
However, for organisms exhibiting sweepstakes reproduction, such as
oysters and other marine organisms
\citep{A04,B94,H82,H94,Eldon2006,Sargsyan2008,E11}, the Kingman
coalescent may not be appropriate, as it only allows binary mergers of
ancestral lineages. Thus, we consider models that allow more than two
lineages to coalesce simultaneously in the gene trees, that is
multiple merger coalescents, also known as $\Lambda$ coalescents
\citep{Pitman1999,S99,DK99}.  The concordance probabilities between
multiple merger gene genealogies and a species tree of two species are
investigated by \cite{ED12}.  

\begin{figure}[htp]
\centering

\begin{tikzpicture}[ultra thick]
\ifx\du\undefined
  \newlength{\du}
  \fi
  \setlength{\du}{2\unitlength}
\draw(-55\du,0\du)--(-35\du,0\du);
\draw(-20\du,0\du)--(-10\du,0\du);
\draw(-10\du,0\du)--(-5\du,15\du);
\draw(-20\du,0\du)--(-15\du,15\du);
\draw(-15\du,15\du)--(-11\du,21.5\du);
\draw(-11\du,21.5\du)--(-15\du,25\du);
\draw(-35\du,0\du)--(-15\du,25\du);
\draw(-55\du,0\du)--(-35\du,25\du)--(-25\du,45\du)--(-20\du,50\du)--(-20\du,65\du);
\draw(0\du,22.5\du)--(-10\du,32.5\du);
\draw(-5\du,45\du)--(-10\du,32.5\du);
\draw(40\du,0\du)--(30\du,0\du);
\draw(15\du,0\du)--(0\du,0\du);
\draw(0\du,0\du)--(-5\du,15\du);
\draw(15\du,0\du)--(10\du,15\du);
\draw(10\du,20\du)--(10\du,15\du);
\draw(10\du,20\du)--(14\du,30\du);
\draw(30\du,0\du)--(14\du,30\du);
\draw(40\du,0\du)--(25\du,30\du)--(5\du,50\du)--(5\du,65\du);
\draw(0\du,22.5\du)--(5\du,35\du);
\draw(-5\du,45\du)--(5\du,35\du);
\draw[thick](-40\du,0\du)--(-39\du,1\du)--(-39\du,2\du)--(-39\du,3\du)--(-38\du,4\du)--(-40\du,5\du)--(-40\du,6\du)--(-39\du,7\du)--(-39\du,8\du)--(-39\du,9\du)--(-39\du,10\du);
\fill(-39\du,10\du)[black]circle(2.5pt);
\fill(-15\du,50\du)[black]circle(2.5pt);
\fill(-5\du,60\du)[black]circle(2.5pt);
\fill(10\du,40\du)[black]circle(2.5pt);
\node at (-45\du,-5\du)[below]{\bf A};
\node at (-15\du,-5\du)[below]{\bf B};
\node at (7.5\du,-5\du)[below]{\bf C};
\node at (35\du,-5\du)[below]{\bf D};
\node at (-50\du,-7\du)[above]{\small a$_1$};
\node at (-45\du,-7\du)[above]{\small a$_2$};
\node at (-40\du,-7\du)[above]{\small a$_3$};
\node at (-15\du,-7\du)[above]{\small b$_1$};
\node at (5\du,-7\du)[above]{\small c$_1$};
\node at (10\du,-7\du)[above]{\small c$_2$};
\node at (35\du,-7\du)[above]{\small d$_1$};
\draw[thick](-39\du,10\du)--(-38\du,12\du)--(-36\du,14\du)--(-36\du,16\du)--(-35\du,18\du)--(-32\du,20\du)--(-32\du,22\du)--(-31\du,24\du)--(-30\du,26\du)--(-29\du,28\du)--(-28\du,30\du)--(-25\du,32\du)--(-24\du,34\du)--(-23\du,36\du)--(-23\du,38\du)--(-22\du,40\du)--(-20\du,42\du)--(-18\du,44\du)--(-18\du,46\du)--(-17\du,48\du)--(-15\du,50\du);
\draw[thick](-15\du,50\du)--(-15\du,51\du)--(-13\du,52\du)--(-12\du,53\du)--(-11\du,54\du)--(-10\du,55\du)--(-9\du,56\du)--(-8\du,57\du)--(-7\du,58\du)--(-7\du,59\du)--(-5\du,60\du);
\draw[thick](10\du,40\du)--(9\du,42\du)--(6\du,44\du)--(5\du,46\du)--(4\du,48\du)--(3\du,50\du)--(0\du,52\du)--(0\du,54\du)--(-1\du,56\du)--(-4\du,58\du)--(-5\du,60\du);
\draw[thick](35\du,0\du)--(34\du,2\du)--(33\du,4\du)--(32\du,6\du)--(31\du,8\du)--(30\du,10\du)--(30\du,12\du)--(29\du,14\du)--(27\du,16\du)--(26\du,18\du)--(26\du,20\du)--(24\du,22\du)--(23\du,24\du)--(23\du,26\du)--(22\du,28\du)--(20\du,30\du);
\draw[thick](20\du,30\du)--(19\du,31\du)--(18\du,32\du)--(17\du,33\du)--(15\du,34\du)--(15\du,35\du)--(14\du,36\du)--(12\du,37\du)--(13\du,38\du)--(10\du,39\du)--(10\du,40\du);
\draw[thick](-15\du,0\du)--(-15\du,1\du)--(-15\du,3\du)--(-14\du,4\du)--(-14\du,5\du)--(-13\du,7\du)--(-11\du,9\du)--(-11\du,10\du)--(-11\du,11\du)--(-12\du,12\du)--(-11\du,13\du)--(-11\du,14\du)--(-10\du,15\du);
\draw[thick](-10\du,15\du)--(-8\du,16\du)--(-6\du,17\du)--(-4\du,18\du)--(-2\du,19\du)--(0\du,20\du);
\draw[thick](0\du,20\du)--(2\du,24\du)--(3\du,26\du)--(4\du,28\du)--(5\du,30\du)--(6\du,32\du)--(8\du,34\du)--(7\du,36\du)--(10\du,38\du)--(10\du,40\du);
\draw[thick](5\du,0\du)--(5\du,1\du)--(5\du,2\du)--(3\du,5\du)--(3\du,6\du)--(2\du,7\du)--(3\du,8\du)--(2\du,10\du)--(1\du,11\du)--(2\du,12\du)--(1\du,13\du)--(1\du,14\du)--(0\du,15\du);
\draw[thick](10\du,0\du)--(10\du,1\du)--(8\du,4\du)--(7\du,5\du)--(8\du,6\du)--(8\du,7\du)--(7\du,8\du)--(7\du,10\du)--(6\du,13\du)--(5\du,14\du)--(5\du,15\du);
\draw[thick](0\du,15\du)--(-2\du,18\du)--(-5\du,19\du)--(-5\du,20\du)--(-7\du,21\du)--(-6\du,22\du)--(-8\du,23\du)--(-9\du,24\du)--(-10\du,25\du)--(-12\du,26\du)--(-12\du,27\du)--(-14\du,28\du)--(-15\du,29\du)--(-15\du,30\du);
\draw[thick](-15\du,30\du)--(-17\du,35\du)--(-16\du,40\du)--(-14\du,45\du)--(-15\du,50\du);
\draw[thick](5\du,15\du)--(5\du,17\du)--(5\du,19\du)--(7\du,21\du)--(6\du,23\du)--(6\du,25\du)--(7\du,27\du)--(7\du,29\du)--(8\du,31\du)--(8\du,33\du)--(10\du,35\du)--(10\du,37\du)--(10\du,40\du);
\draw[thick](-45\du,0\du)--(-44\du,1\du)--(-44\du,2\du)--(-43\du,3\du)--(-42\du,4\du)--(-41\du,5\du)--(-41\du,6\du)--(-40\du,7\du)--(-39\du,8\du)--(-40\du,9\du)--(-39\du,10\du);
\draw[thick](-50\du,0\du)--(-49\du,1\du)--(-48\du,2\du)--(-45\du,3\du)--(-44\du,4\du)--(-45\du,5\du)--(-44\du,6\du)--(-41\du,7\du)--(-40\du,8\du)--(-39\du,9\du)--(-39\du,10\du);
\draw[thick,->](-45\du,30\du)--(-36\du,25\du);
\node at (-45\du,30\du)[above]{\footnotesize species network};
\draw[thick,->](15\du,55\du)--(6\du,45\du);
\node at (15\du,55\du)[right]{\footnotesize gene genealogies};
\draw[thick,->](20\du,50\du)--(11\du,41\du);
\node at (20\du,50\du)[right]{\footnotesize multiple merger};
\draw[thick,->](-30\du,55\du)--(-17\du,51\du);
\draw[thick,->](-30\du,55\du)--(-7\du,60\du);
\node at (-30\du,55\du)[left]{\footnotesize binary mergers};
\draw[thick,->](-50\du,15\du)--(-41\du,11\du);
\node at (-55\du,15\du)[above]{\footnotesize multiple merger};
\end{tikzpicture}
\caption{Example of a multiple merger gene genealogy with topology 
  {\tt (((a$_1$,a$_2$,a$_3$),c$_1$),(b$_1$,c$_2$,d$_1$))} simulated in
  a species network with topology  {\tt
    ((((B,C)s1)h1\#H1,A)s2,(h1\#H1,D)s3)r}, where {\tt H1} is the probability that a lineage has its ancestry from its left parental population.}\label{fig}
\end{figure}
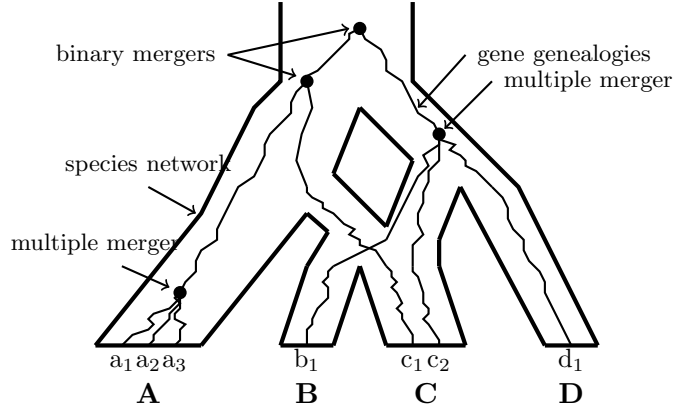

Species trees may also fail to be bifurcating due to either polytomies or hybridization events.
Simulating gene trees from a rooted species network modeling hybridization
is another application of {\tt hybrid-Lambda}. The package {\tt ms}
\citep{Hudson2002ms} can simulate gene trees within a general species
network.  However the input of {\tt ms} is difficult to automate when the
network is sophisticated or generated from other software. Other simulation studies using species networks
have either used a small number of network topologies coded individually  (for example, in {\tt phylonet} \citep{Than2008}
or have assumed that gene trees have evolved on species trees embedded within the species network \citep{Holland2008,Meng2009,Kubatko2009}.  The software {\tt hybrid-Lambda} will help automate simulation studies of hybridization allowing for a large number of species network topologies and allowing gene trees to evolve directly within the network.

\section{DESCRIPTION}

\indent The program input file for {\tt hybrid-Lambda} is a character string that describes
relationships between species. Standard Newick format
\citep{Olsen1990} is used for inputting species trees and outputting
gene trees, whose interior nodes are not labelled. An extended Newick
formatted string \citep{Cardona12008,Huson2010} labels all internal
nodes, and is used for inputting species networks (see
Fig.~\ref{fig}).

\subsection{Parameters}

\indent {\tt Hybrid-Lambda} can use  multiple lineages sampled from each species, then
simulate either a Kingman or a multiple merger ($\Lambda$) coalescent within a given species network.
The coalescent is a continuous-time Markov process, in which times
between coalescent events are independent exponential random variables
with different rates. The rates are determined by a so-called
coalescent parameter in the program that can be input via command
line, or a(n) (extended) Newick formatted input string with specific
coalescent parameters as branch lengths. By default, the Kingman
coalescent is used, for which a population with $b$ lineages sampled has two lineages coalesce at rate $\lambda_{b,2} = \binom{b}{2}$.  One can choose
between two different examples of a $\Lambda$ coalescent.  If the
coalescent parameter is between 0 and 1, then we use $\psi$ for the coalescent parameter, and the rate $\lambda_{bk}$
at which $k$ out of $b$ active ancestral lineages merge is 
\begin{equation}
\label{eqn:psi}
\lambda_{bk}=\binom{b}{k}\psi^k(1-\psi)^{b-k},\quad \psi \in (0,1),
\end{equation}
and if the coalescent parameter is between 1
and 2, then we use $\alpha$ for the coalescent parameter, and the rate of $k$-mergers is 
\begin{equation}
\label{eqn:alpha}
\lambda_{bk}=\binom{b}{k}\frac{B(k-\alpha,b-k+\alpha)}{B(2-\alpha,\alpha)}, \quad \alpha \in (1,2),
\end{equation}
where $B(\cdot,\cdot)$  is the beta function \citep{Schweinsberg2003}.

The program {\tt hybrid-Lambda} assumes that the input network (tree) branch lengths are in coalescent units. However, this is not essential. Coalescent units can be converted through an alternative input file with numbers of generation as branch lengths, then divided by its corresponding effective population sizes.
By default, effective population sizes on all branches are assumed to be equal and unchanged. Users can change this parameter using the command line, or using a(n) (extended) Newick formatted string to specify population sizes on all branches though another input file.

The simulation requires ultrametric species networks, i.e.\  equal lengths of all paths from tip to root. {\tt Hybrid-Lambda} checks the distances in coalescent units between the root and all tip nodes and prints out warning messages if the ultrametric assumption is violated.

\subsection{Output}
\indent {\tt Hybrid-Lambda} outputs simulated gene trees in three different files: one contains gene trees with branch lengths in coalescent units, another uses the number of generations as branch lengths, and the third uses the number of expected mutations as branch lengths. 

Besides outputting gene tree files, {\tt hybrid-Lambda} also provides several functions for analysis purposes:
\begin{itemize}
\item user-defined random seed for simulation,
\item a frequency table of gene tree topologies,
\item a figure of the species network or tree (this function only works when \LaTeX ~or {\tt dot} is installed),
\item when gene trees are simulated from two populations, {\tt hybrid-Lambda} can generate a table of relative frequencies 
of reciprocal monophyly, paraphyly, and polyphyly.
\end{itemize}


\paragraph{Funding:} This work was supported by New Zealand Marsden Fund (SZ and JD), EPSRC (BE).  This work was partly conducted while JD was a Sabbatical Fellow at the National Institute for Mathematical and Biological Synthesis, an Institute sponsored by the National Science Foundation, the U.S. Department of Homeland Security, and the U.S. Department of Agriculture through NSF Award \#EF-0832858, with additional support from The University of Tennessee, Knoxville.

\end{document}